\pdfoutput=1%
\documentclass[a4paper,american,floatfix,pdftex,superscriptaddress,twoside,%
aip,jcp,%
citeautoscript,% leave away for latexdiff
reprint,%final% add final to see real layout, no todonotes anymore, ...
]{revtex4-2}%
\usepackage{amsfonts,amsmath,amssymb}
\usepackage[T1]{fontenc}
\usepackage{graphicx}%
\usepackage[utf8]{inputenc}
\usepackage{hypernat}
\usepackage[todos]{CMImacros} % always turn off darkmode, at least, for submission
\usepackage[displaymath,textmath,graphics]{preview}

\graphicspath{{./figures/}} % define figure directory

\newcommand{\cfeldesy}{\affiliation{Center for Free-Electron Laser Science, Deutsches
      Elektronen-Synchrotron DESY, Notkestraße 85, 22607 Hamburg, Germany}}%
\newcommand{\uhhcui}{\affiliation{The Hamburg Center for Ultrafast Imaging, Universität Hamburg,
      Luruper Chaussee 149, 22761 Hamburg, Germany}}%
\newcommand{\uhhphys}{\affiliation{Department of Physics, Universität Hamburg, Luruper Chaussee 149,
      22761 Hamburg, Germany}}%
\newcommand{\ucl}{\affiliation{Department of Physics and Astronomy, University College London, Gower
      Street, London WC1E 6BT, United Kingdom}}
\newcommand{\ayemail}{\email[Email: ]{andrey.yachmenev@cfel.de}}%
\newcommand{\cmiweb}{\homepage[URL: ]{https://www.controlled-molecule-imaging.org}}%

\newcommand*{\ee}{\ensuremath{\text{\emph{ee}}}\xspace}%

\begin{document}
\title{Detecting handedness of spatially oriented molecules by Coulomb explosion imaging}%
\author{Cem Saribal}\cfeldesy\uhhphys%
\author{Alec Owens}\ucl%
\author{Andrey Yachmenev}\ayemail\cmiweb\cfeldesy\uhhcui%
\author{Jochen Küpper}\cfeldesy\uhhphys\uhhcui%
\date{\today}
\begin{abstract}
   We present a new technique for detecting chirality in the gas phase: Chiral molecules are
   spatially aligned in three-dimensions by a moderately strong elliptically-polarized laser field.
   The momentum distributions of the charged fragments, produced by laser-induced Coulomb explosion,
   show distinct three-dimensional orientation of the enantiomers when the laser polarization
   ellipse is rotated by a non-right angle with respect to the norm vector of the detector plane.
   The resulting velocity-map-image asymmetry is directly connected to the enantiomeric excess and
   to the absolute handedness of molecules. We demonstrated our scheme computationally for camphor
   (C$_{10}$H$_{16}$O), with its methyl-groups as marker fragments, using quantum-mechanical
   simulations geared toward experimentally feasible conditions. Computed sensitivity to
   enantiomeric excess is comparable to other modern chiroptical approaches. The present method can
   be readily optimized for any chiral molecule with an anisotropic polarizability tensor by
   adjusting the polarization state and intensity profile of the laser field.
\end{abstract}
\maketitle

Chiral molecules exist in structural forms known as enantiomers, which are mirror images of one
another that are non-superimposable by translation and rotation. The chemical behavior of molecular
enantiomers can be profoundly different. Particularly in the pharmaceuticals industry, methods to
differentiate between them or to determine the enantiomeric excess (\ee) of a chiral sample are
important. In recent years, there have been considerable advances in gas-phase chiroptical
techniques and a variety of such methods have emerged, for example, using phase-sensitive microwave
spectroscopy~\cite{Patterson:Nature497:475, Domingos:ARPC69:499}, Coulomb explosion imaging with
coincidence detection~\cite{Pitzer:Science341:1096, Herwig:Science342:1084}, photoelectron circular
dichroism (PECD)~\cite{Boewering:PRL86:1187, Lux:ACIE51:5001, Janssen:PCCP16:856,
   Kastner:CPC17:1119, Comby:NatComm9:5212}, chiral-sensitive high-harmonic
generation~\cite{Cireasa:NatPhys11:654, Neufeld:PRX9:031002, Baykusheva:PRX8:031060}, or
attosecond-time-resolved photoionization~\cite{Beaulieu:Science358:1288}. These approaches offer
improved sensitivity and their success is based on exploiting electric-dipole interactions for
chiral discrimination~\cite{Ordonez:PRA98:063428}, producing stronger signals than circular
dichroism from magnetic-dipole interactions.

Coulomb explosion imaging is a powerful and efficient approach to retrieve the instantaneous
absolute structures of complex molecules~\cite{Gagnon:JPB41:215104, Slater:PRA89:011401,
   Burt:JCP148:091102}. Applied to chiral molecules, coincident imaging of fragments emitted from
the chiral center can be used to determine the handedness of their enantiomers, in the conceptually
most straightforward way by coincident detection of all fragments attached to the
stereocenter~\cite{Pitzer:Science341:1096, Herwig:Science342:1084}. For axially chiral molecules, it
has been demonstrated that it is sufficient to only correlate two different fragments, if the
molecules are pre-aligned along their axis of chirality~\cite{Christensen:PRA92:033411}.

For molecular enantiomers placed in a field coupling two molecular dipole moment projections or two
off-diagonal polarizability elements it was demonstrated that they exhibit transient dipole moments
and spatial orientations with opposite sign for the different
enantiomers~\cite{Patterson:Nature497:475, Yachmenev:PRL117:033001, Tutunnikov:JPCL9:1105,
   Yachmenev:PRL123:243202, Milner:PRL122:223201}. Experiments inducing enantiomer-specific
orientation, \eg, probed by Coulomb explosion imaging, were reported, albeit so far with very low
sensitivity to the enantiomeric excess~\cite{Milner:PRL122:223201}.

Here, we explore the effect of spatial three-dimensional (3D) alignment of molecules in Coulomb
explosion imaging in order to sensitively probe the \ee and the handedness of chiral molecules with
it. Using accurate computational procedures, we demonstrate that 3D alignment by an
elliptically-polarized non-resonant field can break the symmetry in a fragments position and
momentum distribution in the detector plane, if the polarization ellipse is tilted by an angle
\degree{0<\beta<90} with respect to the norm vector of the detector. The asymmetry between the
detector's left and right halfs gives access to the \ee\ and handedness of chiral molecules. This
method is more robust than previous Coulomb-explosion-based approaches, \eg, regarding detector
limitations and experimental imperfections. Our theoretical estimates for the sensitivity to the
\emph{ee} are comparable to other modern chiroptical techniques, such as PECD. To further enhance
sensitivity we also explore the effect of one-dimensional (1D) orientation combined with 3D
alignment.

\begin{figure*}
   \includegraphics[width=\linewidth]{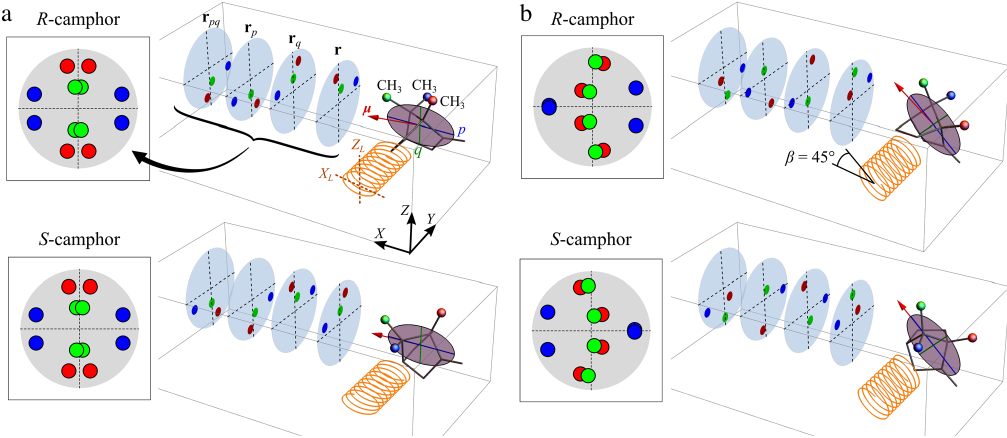}%
   \caption{Sketch of the 3D alignment of the $R$ and $S$ enantiomers of camphor by an
      elliptically-polarized laser field and corresponding projections of its methyl groups onto the
      detector. The most polarizable axes $p$ and $q$ ($\alpha_p>\alpha_q$) of the molecule are
      aligned along the major $Z_\text{L}$ and minor $X_\text{L}$ axes of the elliptical field, and
      the four different projections of the methyl groups onto the plane of the detector correspond
      to the four equivalent molecular orientations in the $(X_\text{L},Z_\text{L})$ plane. (a) When
      either of the $X_\text{L}$ or $Z_\text{L}$ axes is perpendicular to the plane of the detector,
      the sum of the different methyl-group projections look exactly the same for the different
      enantiomers. (b) However, the projections differ when the polarization ellipse is rotated by a
      non-right angle $\beta\neq{}n\cdot\degree{90},~n=0,1,2,\ldots$; see \eqref{eq:coord} for the
      definition of the $\mathbf{r}$ vectors.}
   \label{fig:sketch}
\end{figure*}
\autoref{fig:sketch} illustrates the underlying idea of our approach, which is demonstrated for the
prototypical chiral molecule camphor (C$_{10}$H$_{16}$O). A non-resonant elliptically-polarized
laser field is applied to achieve 3D alignment. The most polarizable axis of the molecule $p$ is
aligned along the major axis $Z_\text{L}$ of the elliptical field and the second most polarizable
axis $q$ along the minor elliptical axis $X_\text{L}$. We chose the three distinct methyl (CH$_3$)
groups in camphor as marker fragments to differentiate between the $R$ and $S$ enantiomers in a
Coulomb-explosion imaging. Their flight directions can be observed experimentally as momentum
distributions of the CH$_3^+$ ions resulting from multiple ionization followed by Coulomb explosion
of the molecule~\cite{Vager:Science244:426, Castilho:RCMSPEC28:1769}.

We assume that two-body dissociation events produce equal initial momenta for CH$_3^+$ fragments at
three different molecular sites. By normalizing the size of the Newton sphere to one, the momenta
distributions are given by the position distributions of the CH$_3$ groups. These methyl-group
distributions in the detector plane are schematically plotted in \autoref{fig:sketch} for the
idealized case of perfect 3D alignment. Fixed in the $X_\text{L}Z_\text{L}$ laser polarization
plane, the molecule orients itself in one of four equally preferred ways, which are related by
\degree{180} rotations about the most polarizable $p$ and $q$ axes of the molecule. Fixing the plane
of elliptical polarization in the $XZ$ laboratory plane, the cartesian coordinates of an atom in the
molecule projected onto the $YZ$ plane of the detector for all four possible spatial molecular
orientations are given by
\begin{equation}
   \begin{aligned}
     \mathbf{r}       &= \left(+y, +(z\cos\beta-x\sin\beta)\right) \\
     \mathbf{r}_p     &= \left(-y, +(z\cos\beta+x\sin\beta)\right) \\
     \mathbf{r}_q     &= \left(-y, -(z\cos\beta+x\sin\beta)\right) \\
     \mathbf{r}_{pq}  &= \left(+y, -(z\cos\beta-x\sin\beta)\right)
  \end{aligned}
  \label{eq:coord}
\end{equation}
where $x,y,z$ denote the cartesian coordinates of an atom in the principal-axis-of-polarizability
frame of the molecule. The subscript indices $p$ and $q$ denote cartesian vectors obtained by
\degree{180} rotations about the respective molecular polarizability axes, which in the case of
perfect 3D alignment coincide with the $Z_\text{L}$ and $X_\text{L}$ axes of the polarization
ellipse. The angle $\beta$ is the angle between the major $Z_\text{L}$ axis of the ellipse and the
norm ($\mathbf{e}_X$) of the detector. It describes the rotation of the polarization ellipse about
the $Y$ axis.

The four different positions $\mathbf{r}$, $\mathbf{r}_p$, $\mathbf{r}_q$ and $\mathbf{r}_{pq}$ in
the plane of the detector are plotted in \autoref{fig:sketch} for the three carbon atoms that belong
to the methyl-groups for $R$ and $S$ camphor. Different enantiomers have opposite signs of the $Y$
component of each position vector in \eqref{eq:coord}. When
$\beta=n\cdot\degree{90},~n=0,1,2,\ldots$ the four different positions in the plane of the detector
for each atom produce an image, which is symmetric with respect to the inversion of both $Y$ and $Z$
axes, as shown in \autoref[a]{fig:sketch}. Since the position vectors for the $R$ and $S$
enantiomers differ only in the sign of the $Y$ coordinate, the resulting projections will look
exactly the same for different enantiomers. However, when $\beta\neq{}n\cdot\degree{90}$ the
symmetry with respect to the inversion of the $Y$ axis in \eqref{eq:coord} will be broken. As a
result, the sums of the four equivalent molecular spatial orientations will exhibit distinctly
different projections on the detector plane for the $R$ and $S$ enantiomers, see
\autoref[b]{fig:sketch}. The detector images of the enantiomers are asymmetric with respect to the
left and right parts and are in fact mirror images of each other for the enantiomers. This allows
for the determination of the \ee\ and the handedness of chiral molecules. Notably, the present
approach does not require coincidence measurements of different fragment species.

To benchmark our scheme we performed quantum-mechanical calculations of the rotational dynamics
of camphor using the accurate variational procedure RichMol~\cite{Owens:JCP148:124102}, which
simulates the rotation-vibration dynamics of molecules in the presence of external fields. The
field-free rotational motion was modelled using the rigid-rotor Hamiltonian with the rotational
constants $A=1446.968977$~MHz, $B=1183.367110$~MHz, and $C=1097.101031$~MHz~\cite{Kisiel:PCCP5:820}.
Simulations of the field-induced time-dependent quantum dynamics employed wave packets built from
superpositions of field-free eigenstates including all rotational states of the molecule with
$J\leq40$, where $J$ is the quantum number of overall angular momentum. Only the vibrational
ground-state was considered, reflecting the conditions in a cold molecular beam. The time-dependent
coefficients were obtained by numerical solution of the time-dependent Schrödinger equation using
the time-discretization method with a time step of $\Delta{t}=10$~fs and a Lanczos-based approach
for the time-evolution operator~\cite{Sidje:TOMS24:130}.

The field interaction potential was represented as a multipole moment expansion of order up to the
polarizability interaction tensor. The dipole moment and polarizability tensor were calculated using
the coupled cluster method CCSD(T) with the augmented correlation-consistent basis set
aug-cc-pVTZ~\cite{Dunning:JCP90:1007, Kendall:JCP96:6796} in the frozen-core approximation. The
calculations were performed at the experimentally determined molecular
geometry~\cite{Kisiel:PCCP5:820} using CFOUR~\cite{CFOUR:2017}.

The long elliptically-polarized laser pulse was represented as
\begin{multline}
   E(t) = E_0 \sqrt{4\log2/(\pi\tau^2)} ~ \exp\left(-4\log 2(t-t_0)^2/\tau^2\right) \\
   \times \left[ (\cos(\omega t)\cos\beta + \frac{1}{\sqrt{3}}\sin(\omega t)\sin\beta)\mathbf{e}_X
   \right. \\
   \left. + (\cos(\omega t)\sin\beta - \frac{1}{\sqrt{3}}\sin(\omega t)\cos\beta)\mathbf{e}_Z
   \right]
\end{multline}
with the parameters $E_0=4\times 10^9$~V/cm, corresponding to a laser peak intensity
$I=6\times 10^{11}~\Wpcmcm$, $\omega=800$~nm, $t_0=440$~ps, and $\tau=250$~ps. The calculations were
performed for $\beta$ angles ranging from 0 to $90^\circ$. For some calculations we added the
interaction between the permanent molecular dipole moment and a static electric field of 1 or
5~kV/cm aligned along the detector norm vector $\mathbf{e}_X$. A hypothetical strong probe pulse,
causing the Coulomb explosion, was applied at a time $t=440$~ps corresponding to the peak intensity
of the alignment field. Idealized simulations were performed at an initial rotational temperature of
$T=0$~K, and for experimentally realistic conditions at $T=0.2$~K. Sub-Kelvin rotational
temperatures can routinely be achieved using carefully optimized supersonic
expansions~\cite{Wang:PRL60:696, Hillenkamp:JCP118:8699, Johny:CPL721:149}, molecular beams coupled
to the electrostatic deflector~\cite{Filsinger:JCP131:064309, Chang:IRPC34:557,
   Trippel:RSI89:096110} or focusers~\cite{Filsinger:PRL100:133003, Ghafur:NatPhys5:289,
   Meerakker:CR112:4828}. Alternatively, helium nanodroplets provide comparably low temperatures of
0.4~K~\cite{Hartmann:Science272:1631} and allow for similar Coulomb explosion imaging experiments of
aligned molecules~\cite{Chatterley:PRL119:073202}, including some large and complex
systems~\cite{Chatterley:NatCommun10:133}. Beyond that, buffer-gas cooled molecular beams provide
molecules in the gas phase at temperatures down to $\ordsim1$~K~\cite{Hutzler:CR112:4803} or using
dilution refrigerators even at $\smaller0.5$~K~\cite{Weinstein:Nature395:148}. Such
buffer-gas-cooled beams were demonstrated for complex molecules~\cite{Patterson:PCCP17:5372} and
recently extended to arbitrarily large molecular systems and
nanoparticles~\cite{Samanta:StructDyn7:024304}.

The degree of 3D alignment is characterized by $\expectation{\cos^2\theta_{p,Z_\text{L}}}=0.84$ and
$\expectation{\cos^2\theta_{q,X_\text{L}}}=0.76$ for $T=0$~K. For a finite initial temperature of
$T=0.2$~K we obtained $\expectation{\cos^2\theta_{p,Z_\text{L}}}=0.64$ and
$\expectation{\cos^2\theta_{q,X_\text{L}}}=0.50$.

The distributions of the methyl-group fragments of camphor in the $YZ$ detector plane were simulated
by computing the probability density distributions of the corresponding carbon atoms using the
rotational wavepackets at the peak of the laser pulse. The total distribution was modelled as a
normalized sum of contributions from the three individual methyl-group carbon atoms with equal
weights. As the recoil axes, we chose vectors along the molecular bonds connecting the carbon atoms
in the methyl groups with the backbone of the molecule. To account for non-axial recoil, the
calculated probability density distributions of the methyl-group carbon atoms were convoluted with a
Gaussian function of a solid angle representing angular displacement from the recoil vector. The
full-width at half maximum (FWHM) parameter of the Gaussian function was chosen at \degree{30},
which is near typical experimental values~\cite{Christensen:PRA94:023410}.

\begin{figure*}
   \includegraphics[width=\linewidth]{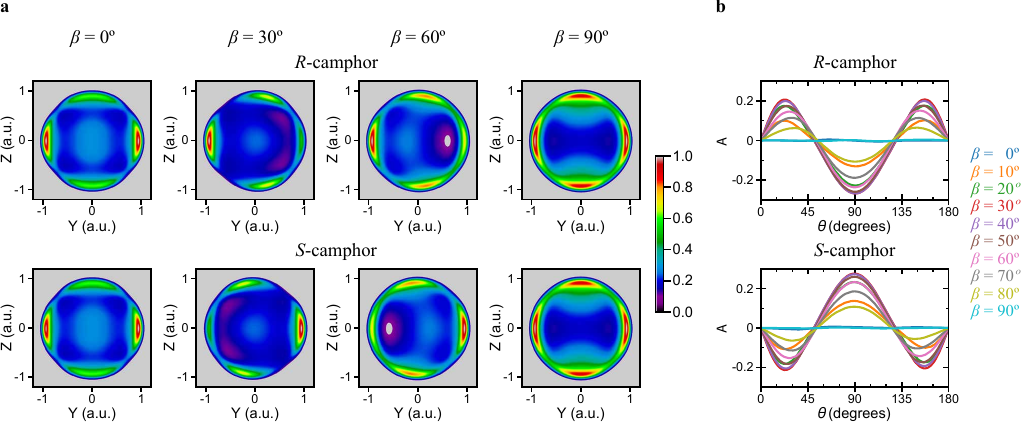}%
   \caption{(a) Computed 2D projections of the averaged probability density distributions of the
      carbon atoms in the methyl groups of $R$ and $S$ camphor at the peak of the alignment field
      and an initial rotational temperature of $T=0$~K. (b) Asymmetry parameter $\mathcal{A}$ as a
      function of the $\theta$ angle. The results are shown for angles $\beta=\degree{0,30,60,90}$
      between the major axis $a$ of the elliptical field and the norm vector of the detector plane.}
   \label{fig:PDF=0}
\end{figure*}
\autoref[a]{fig:PDF=0} shows the calculated 2D projections of the probability density distributions
for the carbon atoms in the methyl groups of $R$ and $S$ camphor for different $\beta$ angles and an
initial rotational temperature of $T=0$~K. As expected, for $\beta=0,\degree{90}$, the 2D
projections are symmetric with respect to inversion of $Y$ and $Z$ axes. Thus, their averages for
the four orientation look identical for the different enantiomers. The 2D density projections become
asymmetric with respect to inversion of the $Y$ axis for intermediate values of the $\beta$ angle.
In \autoref[a]{fig:PDF=0} the results are shown for $\beta=30^\circ$ and $60^\circ$. For different
enantiomers the distributions are exact mirror images of each other in the $YZ$ plane. For racemic
mixtures, the 2D density, and consequently the momentum projections of the methyl-group fragments,
will be symmetric to inversion of the $Y$ axis, and the presence of an asymmetry between the left
and right halfs of the detector will thus indicate the \ee.

To identify the parts of the detector images, which have the largest asymmetry and are therefore
most sensitive to the \ee, we propose to define an asymmetry parameter as a normalized difference
$\mathcal{A}(\theta)=[N_\Omega(\theta)-N_\Omega(-\theta)]/[N_\Omega(\theta)+N_\Omega(-\theta)]$
between sectors in the right and left halfs of the detector. Here, $N_\Omega(\theta)$ is the
intensity in an angular sector of fixed width $\Omega$ at $\theta=0\ldots\degree{180}$, \ie, in the
right half of the detector. Thus, $N_\Omega(-\theta)$ is the corresponding intensity in the left
half of the detector. The asymmetry $\mathcal{A}(\theta)$ is linearly dependent on the \ee: it is
zero for the racemic mixture and attains its maximum value for the pure enantiomer. The asymmetry
$\mathcal{A}(\theta)$ for $\Omega=\degree{30}$ for different $\beta$ is shown in
\autoref[b]{fig:PDF=0}. The largest value of $\mathcal{A}$ for the $R$ and $S$ enantiomers,
respectively, are obtained as $\mathcal{A}\approx0.22$ for $\beta=30\ldots\degree{50}$ and
$\mathcal{A}\approx-0.3$ for $\theta=90^\circ$.

Generally, the asymmetry values $\mathcal{A}$ depend on the molecule, its marker fragments, and
their recoil axes with respect to the alignment plane. In the case of a large number of
indistinguishable fragment groups attached at various molecular sites, e.g. hydrogen
atoms~\cite{Mullins:indole:inprep}, the total probability density will look more isotropic, even for
strong 3D alignment. The degree of angular asymmetry will also be lowered when looking at fragments
dissociating in directions nearly co-planar to either the alignment plane or the plane of detector.

In the present case, there are three indistinguishable CH$_3$ fragments attached at different sites
of camphor. The optimal value of the $\beta$ angle can be thought of as the one that maximizes the
overlap of the 2D probability density distributions of different CH$_3$ fragments. This leads to a
more anisotropic total density distribution and a better contrast with respect to variaton of
$\theta$.

\begin{figure}
   \includegraphics[width=\linewidth]{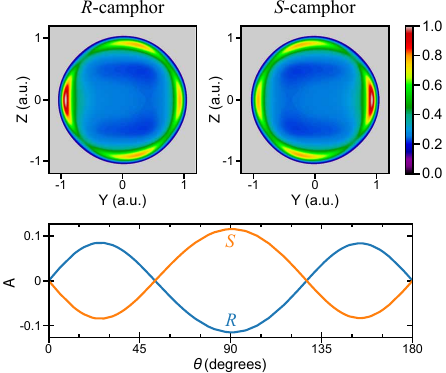}%
   \caption{Computed 2D projections of the averaged probability density distributions for carbon
      atoms in the methyl groups of $R$ and $S$ camphor at the peak of the alignment field and an
      initial rotational temperature of $T=0.2$~K. On the bottom panel, the asymmetry parameter
      $\mathcal{A}$ as a function of the $\theta$ angle for both enantiomers is displayed. The
      results are shown for the optimal value of $\beta=40^\circ$.}
   \label{fig:PDF>0}
\end{figure}

The magnitude of angular asymmetry $\mathcal{A}(\theta)$ also depends on the degree of 3D alignment.
The lower degree of alignment for a 0.2~K sample leads to more diffuse 2D projections of the
probability density distributions and, therefore, to smaller values of asymmetry
$\mathcal{A}(\theta)$. These are plotted in \autoref{fig:PDF>0} for a selected optimal value of
$\beta=40^\circ$. The maximum value of $\mathcal{A}(\theta=\degree{90})=\pm0.1$ at $T=0.2$~K is
decreased by a factor of three as compared to the $T=0$~K results. For higher temperatures close to
1~K, the asymmetry drops further by a factor of 5.3. The loss of asymmetry will vary for different
molecules depending on the density of rotational states as well as their polarizability anisotropy.
The present estimates of the maximum asymmetry for cold ($T\leq 0.4$~K) molecular beams of camphor
are comparable to those achieved in PECD experiments, where the asymmetry is defined as the
normalized difference between the number of electrons emitted by the molecule in the forward and
backward hemispheres relative to the laser beam~\cite{Janssen:PCCP16:856}. Fenchone, for example, a
chiral molecule with a structure similar to the one of camphor, showed an asymmetry value of
$\pm 0.15$ in PECD experiments~\cite{Kastner:CPC17:1119}.
A key advantage of our approach over methods such as PECD or microwave three-wave mixing is access
to the absolute handedness of the \ee. Indeed, the position of the methyl groups with respect to the
plane of 3D alignment is unique for the $R$ and $S$ enantiomers. As a result, the absolute sign of
the left-right asymmetry in the ion momentum distributions can be unambiguously assigned to the
enantiomer's absolute configuration. Notably, in order to predict the absolute sign of the asymmetry
parameter in the axial recoil approximation it is sufficient to know the geometry of the molecule
and its polarizability tensor, where only relative magnitudes of tensor elements matter.

\begin{figure}[t]
   \includegraphics[width=\linewidth]{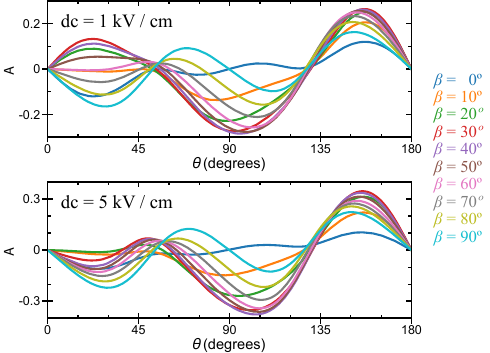}
   \caption{Effect of the dc field on the asymmetry of the 2D projections of the averaged
      probability density distributions for carbon atoms in the methyl groups of $R$-camphor,
      computed at the peak of the alignment field for different values of $\beta$ and an initial
      rotational temperature of $T=0$~K.}
   \label{fig:orient}
\end{figure}

One may consider to increase the degree of asymmetry by rendering the four equivalent alignment
orientations of unequal probability. This can be achieved, for instance, by applying a dc electric
field along the norm vector of the detector plane, known as mixed-field
orientation~\cite{Friedrich:JCP111:6157, Holmegaard:PRL102:023001, Ghafur:NatPhys5:289,
   Nevo:PCCP11:9912}. We calculated the asymmetry $\mathcal{A}(\theta)$ for dc field strengths of 1
and 5~kV/cm at $T=0$~K, shown in \autoref{fig:orient}; note that $\mathcal{A}(S)=-\mathcal{A}(R)$.
As the dc field breaks the symmetry with respect to the inversion of $Y$ and $Z$ axes, although the
simultaneous inversion of both axes is still symmetric, the non-zero asymmetry can be observed even
at $\beta=0,\degree{90}$. The maximal degree of asymmetry increases up to $\pm0.4$ with increasing
dc field strength, the effect however quickly saturates at stronger dc
fields~\cite{Holmegaard:PRL102:023001, Filsinger:JCP131:064309}. The absolute sign of the asymmetry,
defined as the difference between the left and right halfs of the detector, as well as the optimal
values of $\beta$ and $\theta$ remain the same as for pure alignment. This is rationalized by the
fact that the mixed-field orientation in camphor still allows for two of the four orientations
producing the effect of 3D alignment with 1D orientation~\cite{Thesing:JCP146:244304}. The
mixed-field orientation effect however can only be achieved for polar molecules with a non-vanishing
projection of the dipole moment onto the $pq$ plane of the most polarizable axes.

In conclusion, we demonstrated a novel and robust approach for detecting chirality based on the
Coulomb explosion imaging of 3D aligned molecules. The method employs elliptically polarized
non-resonant laser pulses in a standard setup as is typically used for studying molecular
alignment~\cite{Stapelfeldt:RMP75:543, Trippel:MP111:1738}. The chirality is revealed by an
asymmetry in the 2D projections of ion momentum distributions. This paves the way to the sensitive
analytic use of Coulomb explosion imaging for detecting the \ee\ with a sensitivity comparable to
PECD~\cite{Janssen:PCCP16:856, Nahon:PCCP18:12696, Kastner:CPC17:1119}. Any molecule with three
different principal polarizability components can be investigated this way. We note that, different
from PECD, our technique requires strong alignment that is typically achieved by utilizing cold
molecular beams.

Although we found that for camphor the methyl-group fragments deliver sufficient asymmetry, these
fragments could possibly exhibit larger non-axial recoil velocities not fit by the
Gaussian-distribution model assumed. This would result in additional smearing effects on the
structures in the ion momentum distributions. Thus, the present approach is best suited for chiral
molecules with nearly-axially-recoiling leaving groups, but could be extended further through a more
general analysis based on time-resolved measurements~\cite{Mullins:indole:inprep}.

When compared to existing coincidence Coulomb explosion imaging techniques, our approach can
distinguish between the left- and right-handed enantiomers without correlated detection of multiple
different fragments~\cite{Pitzer:Science341:1096, Herwig:Science342:1084, Christensen:PRA92:033411}.
This enables much faster data acquisition, which is highly advantageous for ultrafast time-resolved
studies. For the present method, the asymmetry signal quickly declines with the beam temperature,
with the efficiency similar to coincident imaging at $\ordsim1$~K. The advantage however is that, in
principle, only one fragment type is necessary to detect chirality and handedness, as opposed to
standard methods demanding up to five different fragments. The external fields can be further
optimized to improve the sensitivity. In particular, we demonstrated that mixed-field orientation
can be exploited to enhance the asymmetry in the ion momentum distributions and thus the method's
\ee sensitivity. The approach could be combined with PECD in ion-electron coincidence
measurements~\cite{Ullrich:RPP66:1463} to extract the \ee from the photo-electron distributions
together with the handedness obtained from the ion momentum distributions.

This work has been supported by the Deutsche Forschungsgemeinschaft (DFG) through the priority
program ``Quantum Dynamics in Tailored Intense Fields'' (QUTIF, SPP1840, KU~1527/3, YA~610/1) and
the cluster of excellence ``Advanced Imaging of Matter'' (AIM, EXC~2056, ID~390715994).

The data that support the findings of this study are available from the corresponding author upon
request.

\bibliography{string,cmi}
\end{document}